\documentclass{ws-procs9x6}

\begin{document}

\def\GeV{{\rm GeV}}

\title{Importance of a Measurement of $F_L(x,Q^2)$ at HERA}

\author{R.~S. THORNE\footnote{\uppercase{R}oyal \uppercase{S}ociety 
\uppercase{U}niversity \uppercase{R}esearch \uppercase{F}ellow.}}

\address{Cavendish Laboratory, University of Cambridge, 
J.~J. Thomson Avenue,\\ Cambridge, CB3 0HE\\ 
E-mail: thorne@hep.phy.cam.ac.uk}

\maketitle

\abstracts{I investigate what a direct measurement of the longitudinal 
structure function 
$F_L(x,Q^2)$ could teach us about the structure of the 
proton and the best way in which to use perturbative QCD for structure 
functions. I assume HERA running at a lowered beam energy for approximately
4-5 months and examine how well the measurement could distinguish between 
different theoretical approaches. I conclude that such a measurement would 
provide useful information on how to calculate structure functions and parton 
distributions at small $x$.}

\section{Introduction}

It would be vital to have a real accurate measurement of 
$F_L(x,Q^2)$ at HERA since this gives an independent test
of the gluon distribution at low $x$ to accompany that determined
from $d F_2(x,Q^2)/ d \ln Q^2$ \cite{CTEQ6}--\cite{H1fit}. 
At present the fits to $F_2(x,Q^2)$ 
at low $x$ are reasonably good, but 
the gluon is free to vary in order to make them as successful as possible. 
It is essential to have a cross-check. (It is important to note that
$F_L(x,Q^2)$ is a much better discriminator of the gluon 
distribution, and/or of different theories, for given $F_2(x,Q^2)$ 
than the charm contribution. 
$F^c_2(x,Q^2)$ is constrained to evolve in exactly the 
same way as $F^{tot}_2(x,Q^2)$ (with appropriate charge weighting)
for $W^2 \gg m_c^2$, so is hardly independent.
At lower $W^2$ the supression is determined mainly by kinematics.)
%Also, at lower $W^2$
%the evolution of $F^c_2(x,Q^2)$ is suppressed roughly by a factor of 
%$v= 1-(4m_c^2 z)/(Q^2 (1-z))$
%(the velocity of the heavy quark in the centre-of-mass frame)
%and by the limit of integration in the convolution being 
%$\xi=x(1+4m_c^2/Q^2)$ rather than $x$. 
%Approximately the same amount of suppression seems to exist from 
%order-to-order 
%and when including resummations. Hence $F_2^c(x,Q^2)$ does 
%not distinguish that well between different approaches.
Currently there is a consistency check on the relationship between 
$F_2(x,Q^2)$ and $F_L(x,Q^2)$ at high $y$ since both contribute
to the total cross-section measured at HERA. Hence, there are effective 
``determinations'' of $F_L(x,Q^2)$ obtained by extrapolating
to high $y$ using either NLO perturbative 
QCD or using $(d \sigma/d\ln y)_{Q^2}$ whilst making assumptions 
about $(d F_2(x,Q^2)/d\ln y)_{Q^2}$ \cite{H1FL}. This is a   
good consistency test of a given theory, usually NLO QCD, and
could show up major flaws. However, it relies on small differences between
two large quantities, so its accuracy is limited.
Also, for an extraction of $F_L(x,Q^2)$ it has model-dependent 
uncertainties which are difficult to quantify fully \cite{ThorneFL}. 
A real measurement would be a much more direct test of the success of 
different theories in QCD.

%\begin{figure}[ht]
%\centerline{\epsfxsize=4.1in\epsfbox{h1fldat.ps}}   
%\caption{The consistency check of $F_L(x,Q^2)$ extracted versus
%$F_L(x,Q^2)$ predicted.\label{h1fldat}}
%\end{figure}

\begin{figure}[ht]
\vspace{-2.2cm}
\centerline{\epsfxsize=4.0in\epsfbox{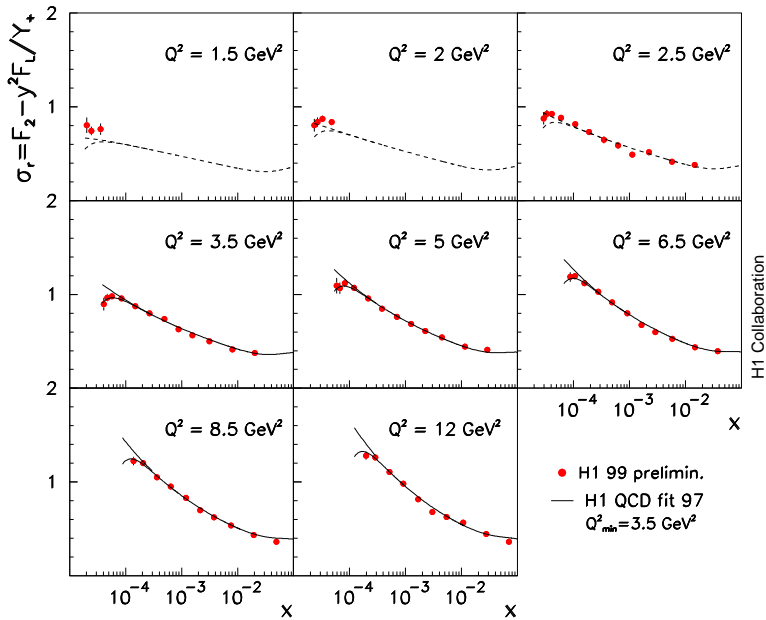}}   
\vspace{-0.5cm}
\caption{The NLO consistency check of  $F_L(x,Q^2)$ for the 
H1 fit.\label{h1flsig}
\vspace{-0.3cm}}
\end{figure}

\begin{figure}[ht]
\centerline{\hspace{-1cm}\epsfxsize=2.4in\epsfbox{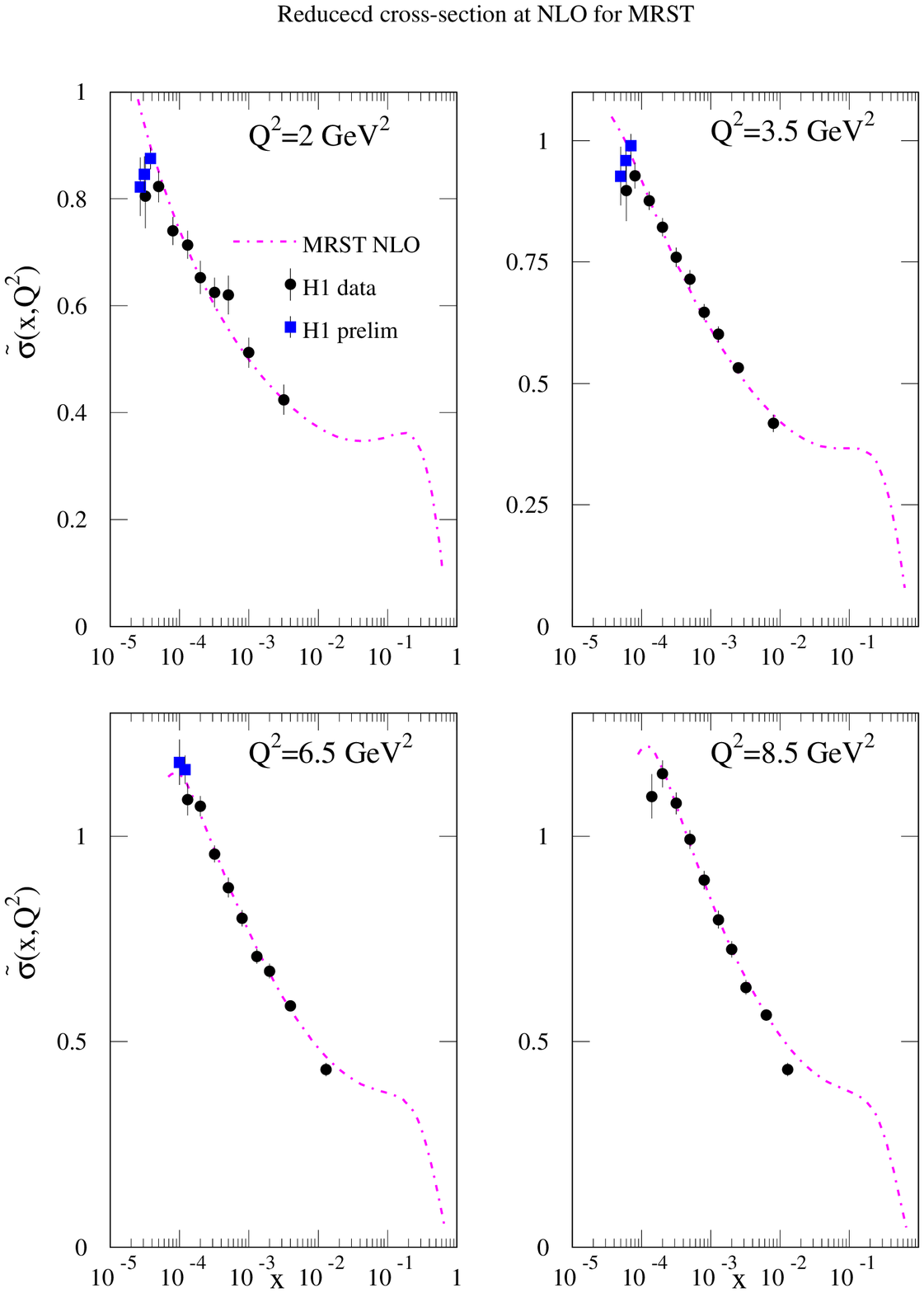}
\hspace{-0.3cm}\epsfxsize=2.4in\epsfbox{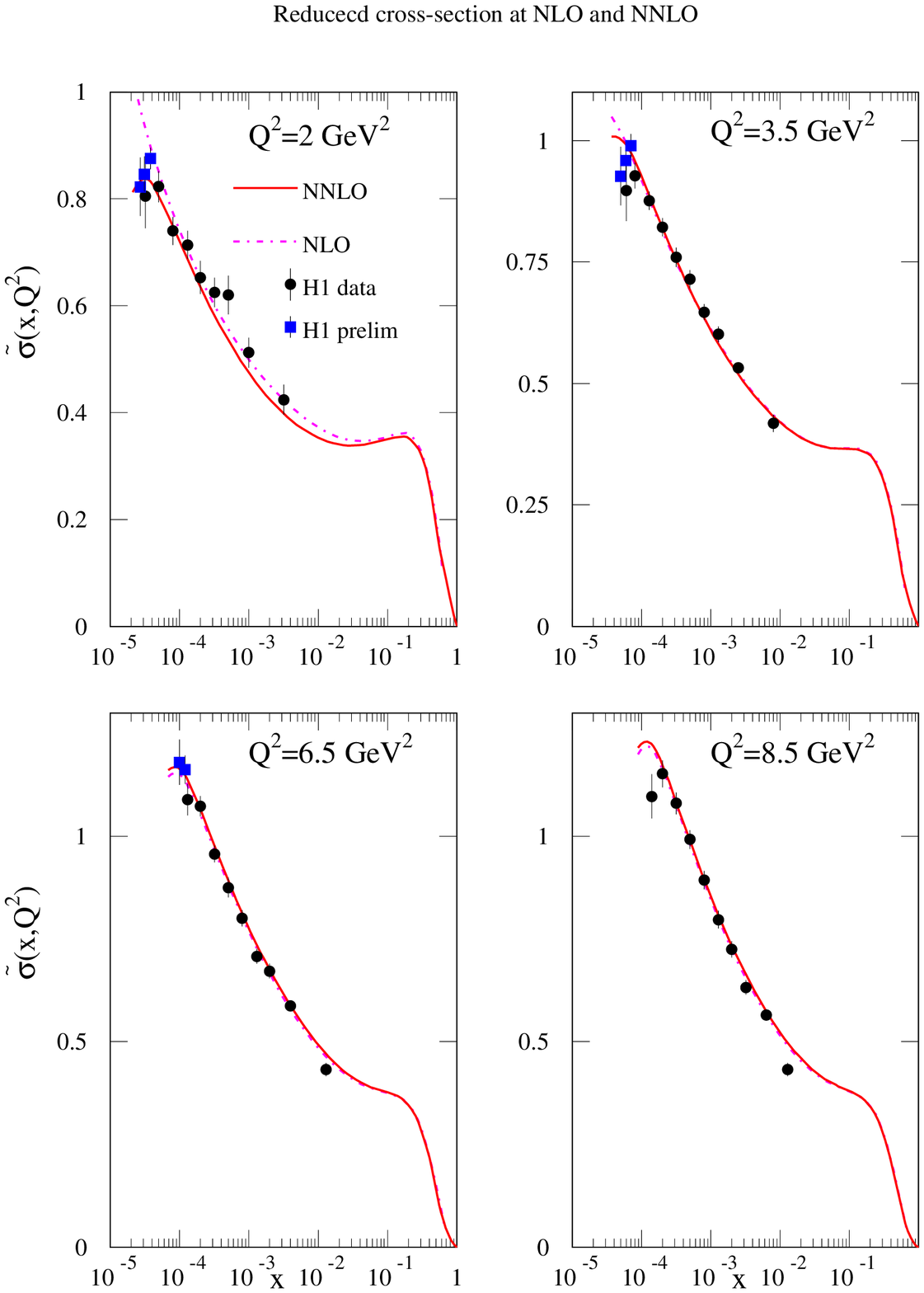}}   
\caption{The NLO consistency check of  $F_L(x,Q^2)$ for the 
H1 fit (left). The consistency check of  $F_L(x,Q^2)$ for the 
NLO and NNLO MRST fits (right).\label{sigrednlo}}
\end{figure}

The consistency check of $F_L(x,Q^2)$ extracted by H1 versus
$F_L(x,Q^2)$ predicted from their QCD fit \cite{H1fit} is shown in Figure
9 in the second of \cite{H1FL}. However, due to the 
potentially large, strongly correlated, 
model-dependent errors in the ``measured'' $F_L(x,Q^2)$
it is far more revealing to see plots like Figure~\ref{h1flsig} \cite{H1FL}.
The turn-over in $\tilde \sigma(x,Q^2)=F_2(x,Q^2)
-y^2/(1+(1-y)^2)F_L(x,Q^2)$ is clearly matched by the 
$F_L(x,Q^2)$ contribution. However, the same consistency check for the fit 
of $\tilde \sigma(x,Q^2)$ 
for MRST partons at NLO fails 
at the lower $Q^2$ values, as seen on the left-hand side of
Figure~\ref{sigrednlo}. This is because of the 
different gluon obtained from a full global fit. Hence, the consistency 
check is not universally successful at NLO.\footnote{Additionally, 
Alekhin performed fits to DIS data, using the 
reduced cross-section for HERA data, and allowed higher-twist 
corrections to be determined phenomenologically.
He found an unambiguous positive 
correction for $F_L(x,Q^2)$, i.e. the consistency 
check fails for the purely perturbative fit \cite{Alekhin}.} 

Hence, many current NLO global fits show problems 
regarding $F_L(x,Q^2)$ at high $y$.
In general they provide a good fit to HERA data, but 
there are some problems in $d F_2/d \ln Q^2$, e.g. Figure~17 in 
\cite{LP03}. However, standard perturbation theory 
 is not necessarily reliable in general because of increasing 
logs at higher orders, e.g. at small $x$
\begin{equation}
P^1_{qg} \sim \alpha_S(\mu^2) \qquad P^2_{qg} 
\sim \frac{\alpha^2_s(\mu^2)}{x} \qquad P^n_{qg} 
\sim \frac{\alpha^n_s(\mu^2)\ln^{n-2}(1/x)}{x}
\end{equation}
\noindent and similarly
\begin{equation}
C^1_{Lg} \sim \alpha_S(\mu^2) \qquad  C^2_{Lg} 
\sim \frac{\alpha_s(\mu^2)}{x}\qquad C^n_{Lg} 
\sim \frac{\alpha^n_s(\mu^2)\ln^{n-2}(1/x)}{x},
\end{equation}
and hence enhancements at higher orders are possible.

%\begin{figure}[ht]
%\centerline{\epsfxsize=4.1in\epsfbox{karlnew1.eps}}   
%\caption{Comparison of MRST(2001) $F_2(x,Q^2)$ with HERA,
%NMC and E665 data.\label{karlnew1}}
%\end{figure}

\begin{figure}[ht]
\centerline{\hspace{-1cm}\epsfxsize=2.4in\epsfbox{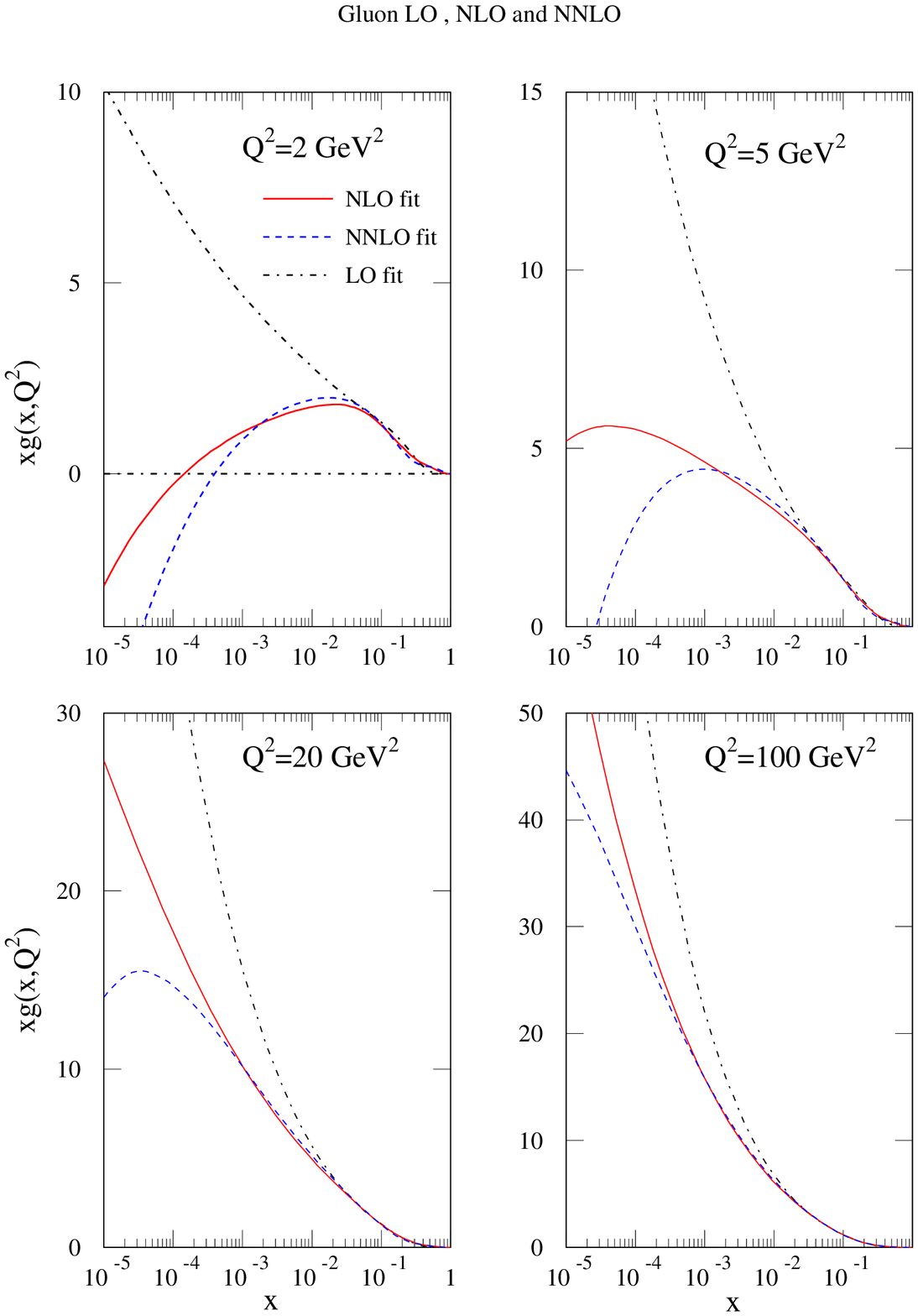}
\hspace{-0.3cm}\epsfxsize=2.4in\epsfbox{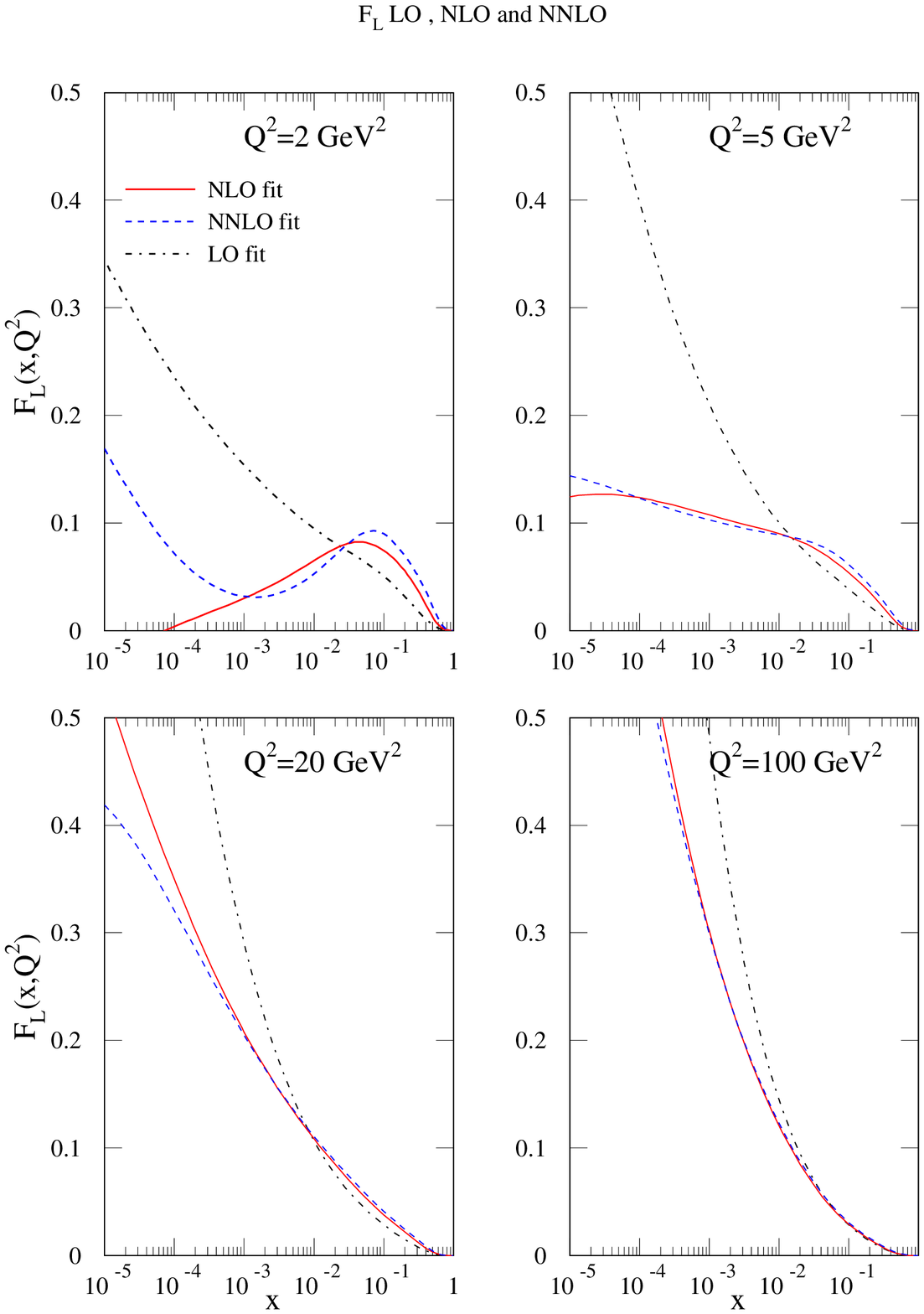}}   
\caption{The gluon extracted from the global fit at LO, NLO and 
NNLO (left). $F_L(x,Q^2)$ predicted from the global fit at LO, 
NLO and NNLO (right).\label{gluewide05}}
\end{figure}

However, we can already see precisely what happens at NNLO.
The splitting functions have been calculated at NNLO \cite{NNLOs}, 
and recently 
the coefficient functions for $F_L(x,Q^2)$ have been finished \cite{NNLOfl}.
The gluon extracted from the MRST global fit at LO, NLO and 
NNLO is shown in Figure~\ref{gluewide05}. 
Additional positive small-$x$ contributions in $P_{qg}$
at each order lead to a smaller low-$x$ gluon at each 
order.\footnote{This conclusion relies on a correct application of flavour 
thresholds in a General Variable Flavour Number Scheme at NNLO \cite{VFNS},
not present in earlier approximate NNLO MRST fits. 
The correct treatment of flavour is particularly important at NNLO
because discontinuities in unphysical quantities appear at this order.}

%\begin{figure}[ht]
%\centerline{\epsfxsize=4.1in\epsfbox{flcoeffg.ps}}   
%\caption{The NNLO 
%$O(\alpha_s^3)$ longitudinal 
%coefficient function $C^3_{Lg}(x)$.\label{flcoeffg}}
%\end{figure}

The NNLO $O(\alpha_s^3)$ longitudinal 
coefficient function $C^3_{Lg}(x)$ given by
\begin{equation}
C^3_{Lg}(x) = n_f \biggl(\frac{\alpha_S}{4\pi}\biggr)^3
\biggl(\frac{409.5\ln(1/x)}{x} -\frac{2044.7}{x}-\cdots\biggr).
\end{equation} 
There is clearly a significant positive contribution at small $x$, and 
this counters the decrease in small-$x$ gluon. 
$F_L(x,Q^2)$ predicted from the global fit at LO, NLO and NNLO is shown in  
Figure~\ref{gluewide05}. The NNLO coefficient function more than 
compensates for the decrease in the NNLO gluon.

%\begin{figure}[ht]
%\centerline{\epsfxsize=4.1in\epsfbox{flnnlowide05.ps}}   
%\caption{$F_L(x,Q^2)$ predicted from the global fit at LO, 
%NLO and NNLO.\label{flnnlowide05}}
%\end{figure}

Without considering the high-$y$ HERA data, the NNLO fit is not much 
better than NLO fit, though it is a slight improvement \cite{MRST04}.   
However, the NNLO contribution to $F_L(x,Q^2)$ largely solves the previous 
high-$y$ problem with $\tilde \sigma(x,Q^2)$, as seen on the right-hand 
side of 
Figure~\ref{sigrednlo}.\footnote{The high-$y$ fit would fail 
with gluons that are positive 
at small $x$ and $Q^2$ -- $F_L(x,Q^2)$ 
would be too big and the turnover too great.}
But these data are not very precise, the effective error on 
$F_L(x,Q^2)$ being $\sim 30-40\%$. It is of real importance to have some 
accurate measurement of $F_L(x,Q^2)$ at small $x$.

%\begin{figure}[ht]
%\centerline{\epsfxsize=4.1in\epsfbox{sigred.ps}}   
%\caption{The NLO consistency check of  $F_L(x,Q^2)$ for the 
%NNLO MRST fit.\label{sigred}}
%\end{figure}

HERA has proposed some running at lower beam energy before finishing
in order to make a direct measurement of $F_L(x,Q^2)$.
The expectation is to measure data from $Q^2 = 5-40 \GeV^2$ and
$x=0.0001-0.003$ with a typical error of at best $12-15 \%$ \cite{Klein}.
How important would this be in distinguishing between different
theoretical approaches to structure functions?
There has been a study by ZEUS \cite{Gwenlandis} on the impact of such data on
the accuracy with which $g(x,Q^2)$ is determined if $F_L(x,Q^2)$ is
roughly as expected from a NLO fit. There is a significant although not
enormous improvement in the gluon uncertainty.
However, this is not, in my view, the most interesting question. 
Rather, it is important to see if the 
potential measurement could tell apart different theoretical treatments, e.g.
whether we need go beyond the standard fixed-order perturbation theory
approach. There has also been a study of this by ZEUS \cite{GwenlanHLHC}, 
with {\it extreme} theoretical predictions, 
and the discriminating power is obvious. However, in this case the 
extremes are based on unrealistic models (out-of-date partons and partons from 
one order used with coefficient functions from another).
Furthermore, all data points are assumed to line up, i.e. the $\chi^2$
for the correct theory would be $0$. A more sophisticated approach is needed. 
   
\section{Test of Theoretical Models}

\begin{figure}[ht]
\centerline{\hspace{-1cm}\epsfxsize=2.1in\epsfbox{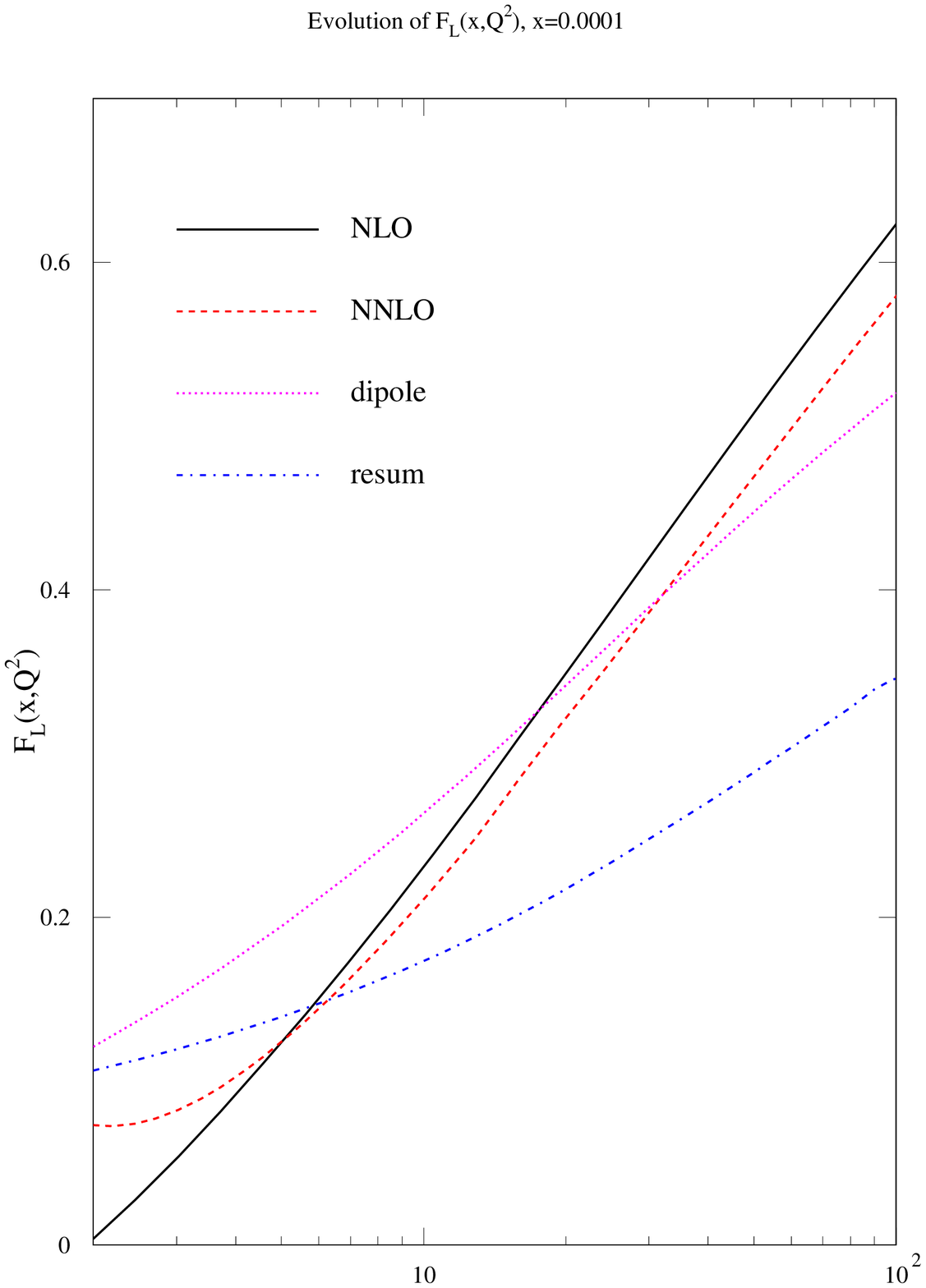}
\hspace{0.1cm}\epsfxsize=2.1in\epsfbox{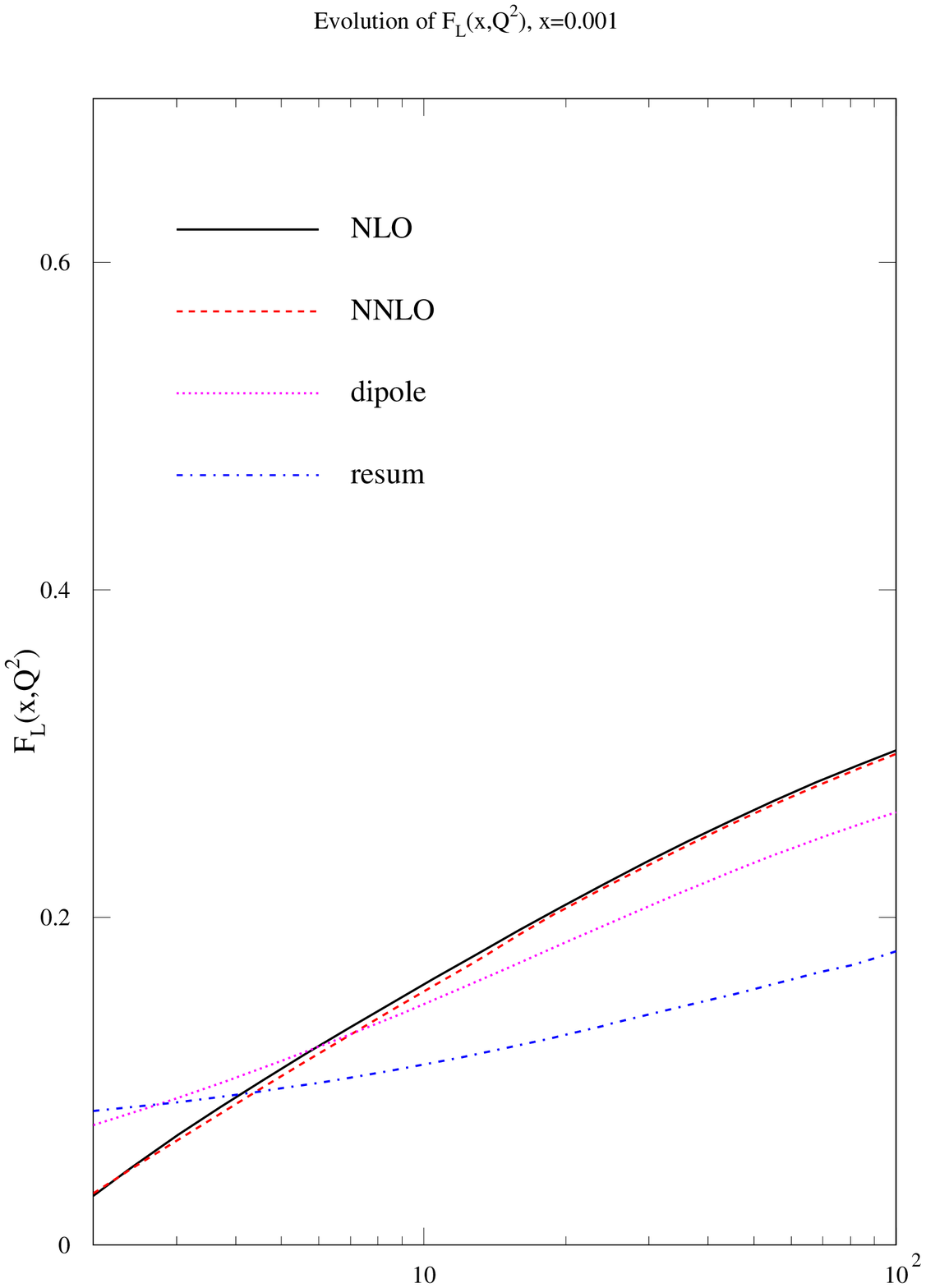}}   
\caption{Evolution of various predictions for $F_L(x,Q^2)$
at $x=0.0001$ (left) and $x=0.001$ (right).\label{flx0001}}
\end{figure}

\begin{figure}[ht]
\centerline{\hspace{-1cm}\epsfxsize=2.4in\epsfbox{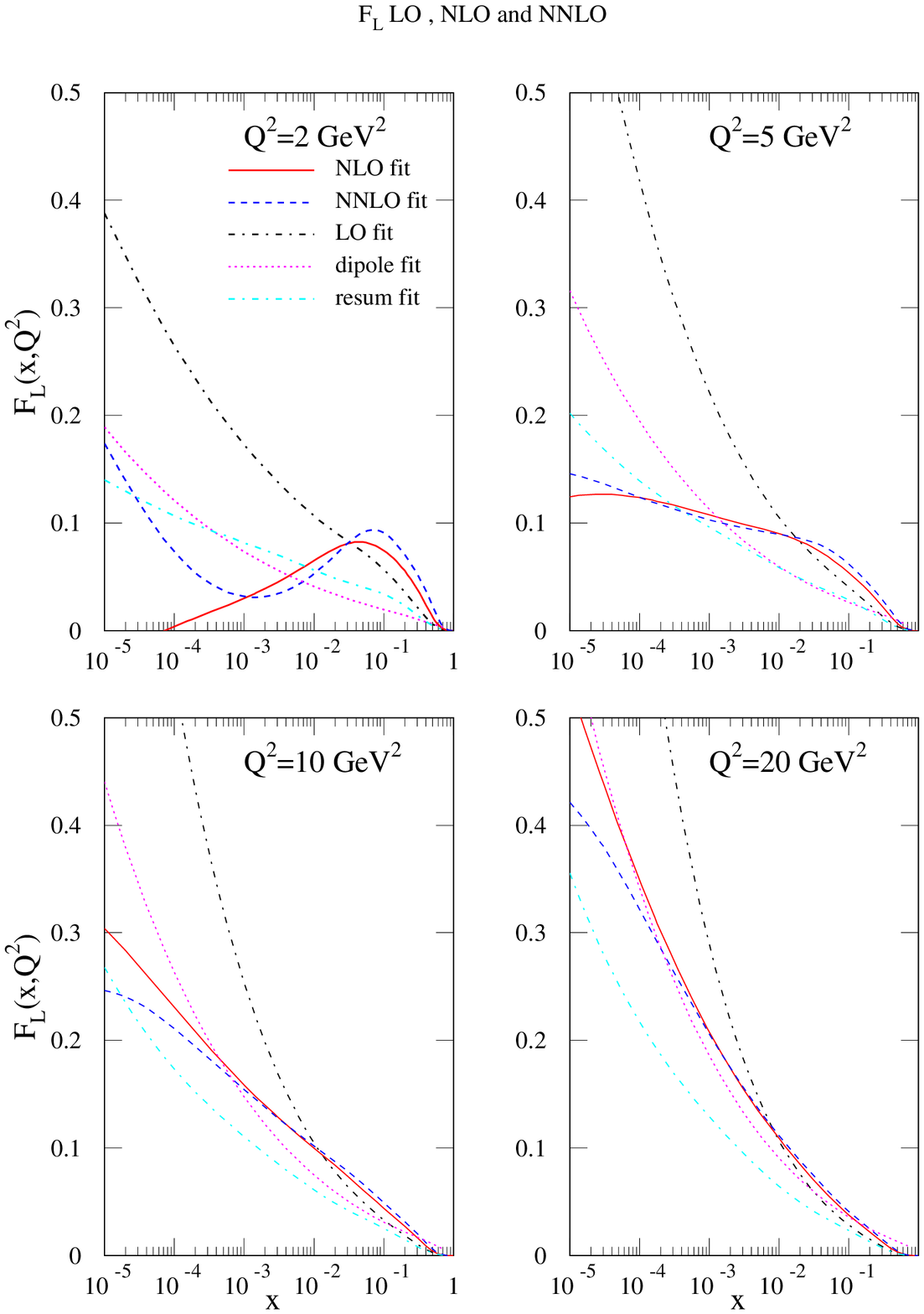}
\hspace{-0.3cm}\epsfxsize=2.4in\epsfbox{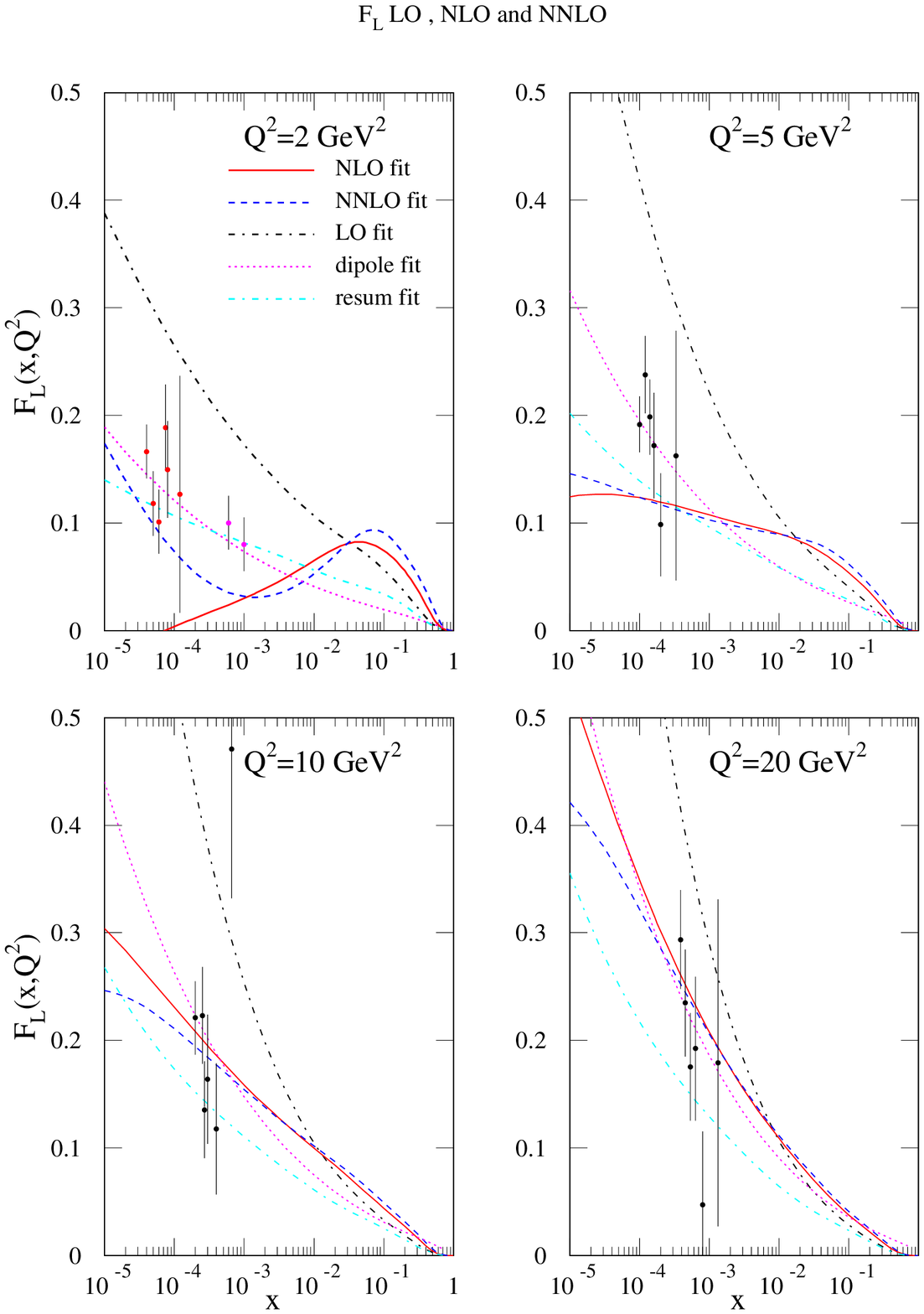}}   
\caption{$F_L(x,Q^2)$ predicted from the global fit at LO, 
NLO and NNLO, from a fit which performs a double resummation 
of leading $\ln(1/x)$ and $\beta_0$ terms, and from a dipole 
model type fit (left). Comparison of data to various predictions. 
Also show are points at $Q^2=2\GeV^2$ that might have been measured at 
HERA III and for the two highest $x$ values 
might be at eRHIC  (right).\label{flnnloringnd}}
\end{figure}

I consider a variety of more plausible theoretical variations. 
A fit that performs a double resummation 
of leading $\ln(1/x)$ and $\beta_0$ terms leads to a better fit
to small-$x$ data than a conventional perturbative fit \cite{resum}.
The resummation also seems to stabilize $F_L(x,Q^2)$ at small $x$ 
and $Q^2$. The fit has some problems at higher $x$ (particularly for 
Drell-Yan data),
and NLO contributions to resummation are needed for 
precision \cite{resumcdw}, hence the 
prediction is somewhat approximate, but it has the correct 
trend.\footnote{Similar results would be likely from the approaches in 
\cite{ccss,abf} since the resummations, though different in detail, have the
same qualitative features.} 
Alternatively, a dipole-motivated fit \cite{gbw}--\cite{Kow} 
contains higher terms in $\ln(1/x)$ and
higher twists. It also guarantees reasonable behaviour for $F_L(x,Q^2)$
at low $Q^2$ due to the form of wavefunction. In a quantitative comparison 
I use my own dipole-motivated fit \cite{RSTdipole} in order to
avoid problems in the heavy flavour treatment in some other approaches. 
The evolution of various predictions for $F_L(x,Q^2)$
at $x=0.0001$ and $x=0.001$, is seen in Figure~\ref{flx0001}. 
The resummation and dipole 
predictions  are behaving sensibly at low $Q^2$. The NLO prediction is 
becoming negative at the lowest values, while the
NNLO prediction is becoming flat at $Q^2 \sim 2 \GeV^2$ for $x=0.0001$. 
It has a 
slight turn up at even smaller $x$, implying the necessity for even 
further corrections. 
The results are shown for various values of $Q^2$ on the left-hand side of 
Figure~\ref{flnnloringnd}. 
They suggest that a measurement of 
$F_L(x,Q^2)$ over as wide a range of 
$x$ and $Q^2$ as possible would be very useful.

%\begin{figure}[ht]
%\centerline{\epsfxsize=4.1in\epsfbox{flnnlodat.ps}}   
%\caption{Proposed measurements of $F_L(x,Q^2)$ by H1 collaboration.
%\label{flnnlodat}}
%\end{figure}

%\begin{figure}[ht]
%\centerline{\epsfxsize=4.1in\epsfbox{claireerr.ps}}   
%\caption{Improvement in uncertainty in gluon density from $F_L(x,Q^2)$ 
%measurement.
%\label{claireerr}}
%\end{figure}

%\begin{figure}[ht]
%\centerline{\epsfxsize=4.1in\epsfbox{clairecomp.ps}}   
%\caption{Attempt to fit extreme prediction for $F_L(x,Q^2)$. 
%\label{clairecomp}}
%\end{figure}

In particular, the dipole fit produces a rather different shape and size 
prediction for $F_L(x,Q^2)$ from that at NLO and NNLO. Hence I 
generate a set of data based on the central dipole prediction but
with a random scatter ($\chi^2 =20/18$ for the dipole prediction).
The comparison of the pseudo-data to other predictions is 
shown on the right-hand side of Figure~\ref{flnnloringnd}, where I  
also show points at $Q^2=2\GeV^2$ that might have been measured at 
HERA III and might be at eRHIC \cite{Caldwell}. 
Points at $40 \GeV^2$ are not as useful, as the errors are bigger and 
the theoretical curves are converging. From Figure~\ref{flnnloringnd} 
it is clear that there is some reasonable differentiating power,  
but this is comparing the central predictions for a given theoretical 
framework only. We must also consider the uncertainties.

%\begin{figure}[ht]
%\centerline{\epsfxsize=4.1in\epsfbox{flnnloring.ps}}   
%\caption{Comparison of data to various predictions. 
%Also show points at $Q^2=2\GeV^2$ which might have been measured at 
%HERA III -- red and might be at eRHIC -- pink. 
%\label{flnnloring}}
%\end{figure}

\begin{figure}[ht]
\centerline{\hspace{-1cm}\epsfxsize=2.4in\epsfbox{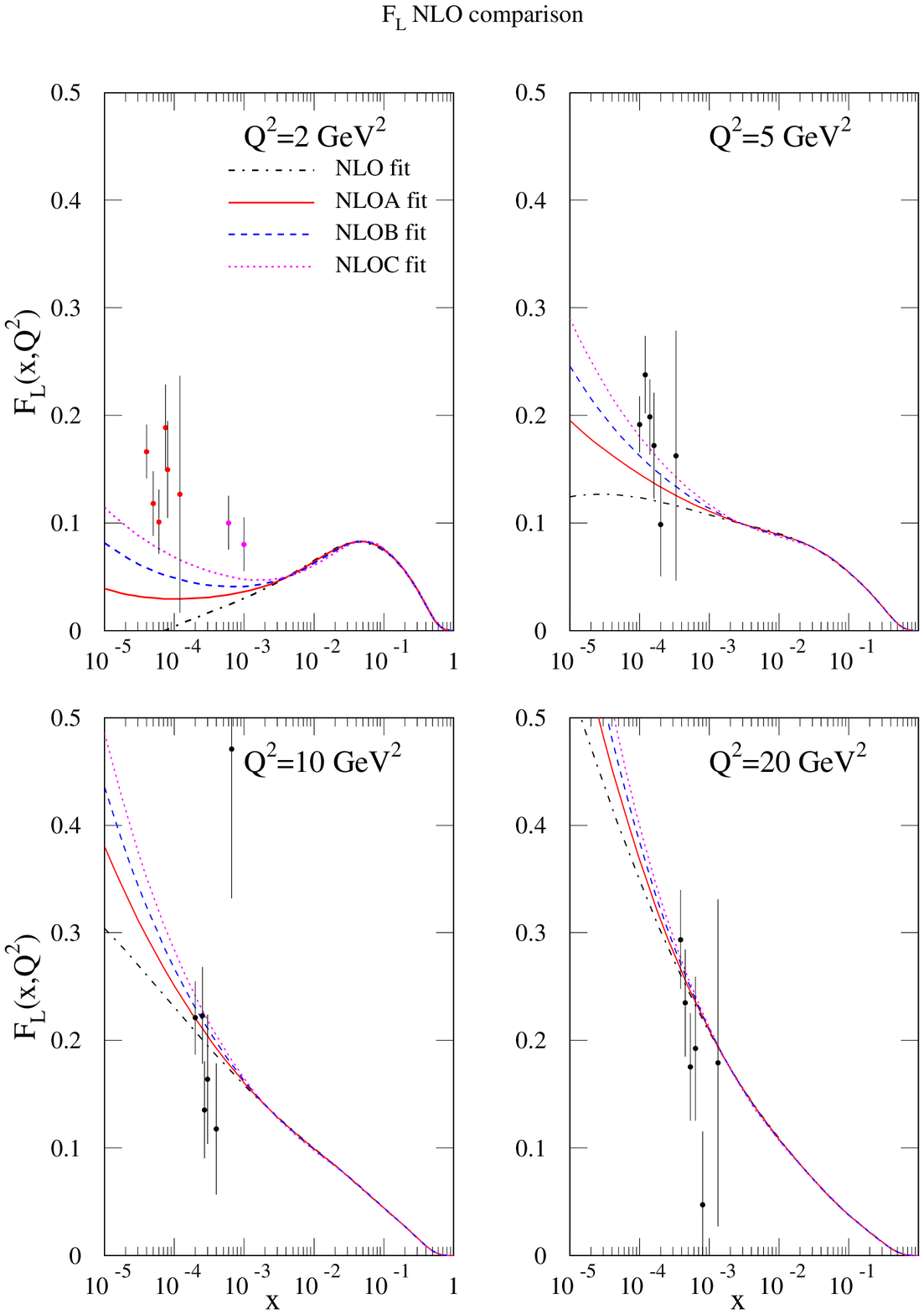}
\hspace{-0.3cm}\epsfxsize=2.4in\epsfbox{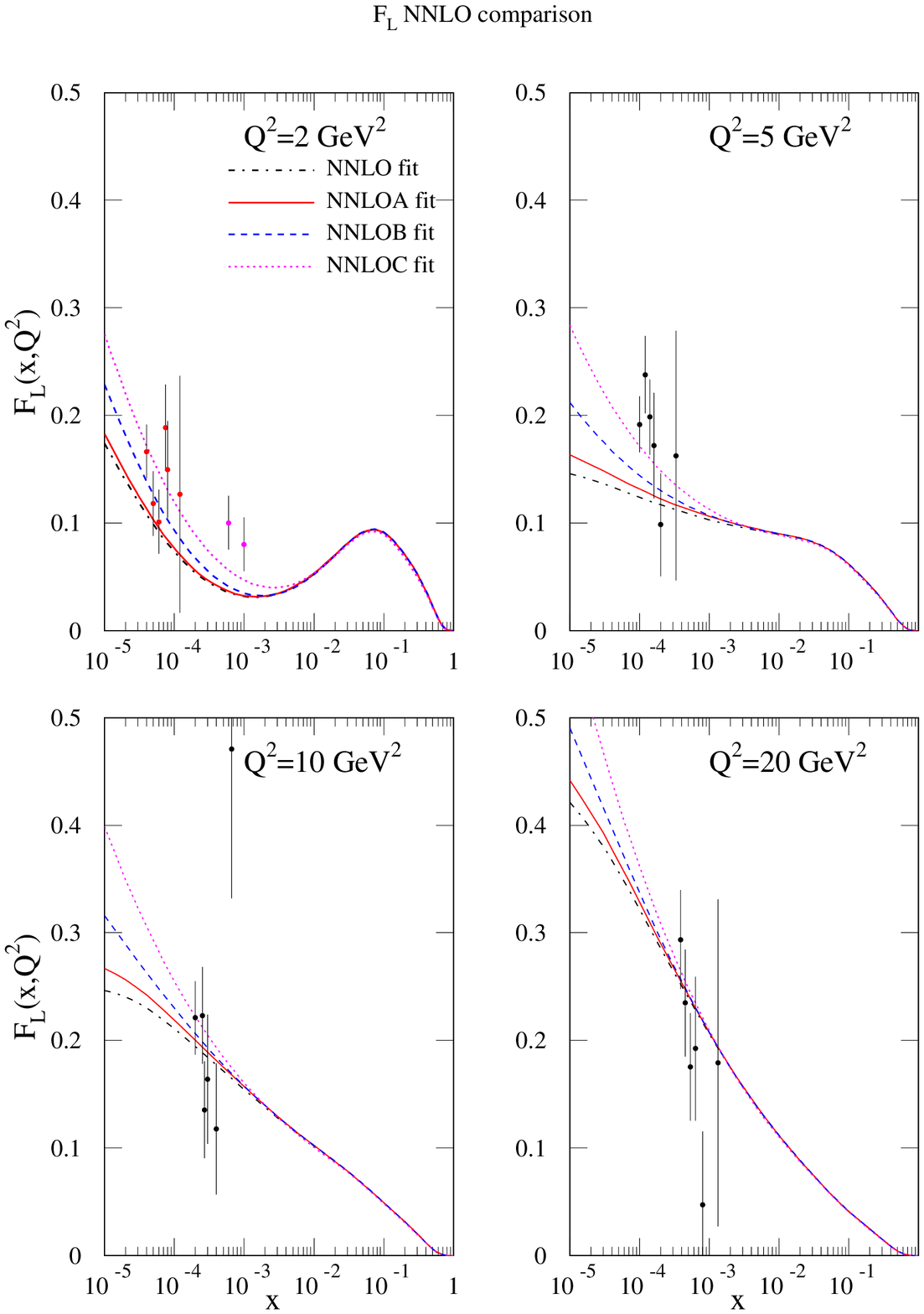}}   
\caption{Comparison at NLO (left) and at NNLO (right) as the 
weight of $F_L(x,Q^2)$ {\it data} is increased in the fit. 
\label{flnlocomp}}
\end{figure}

This is shown in Figure~\ref{flnlocomp}, where
the the left-hand side shows the comparison at NLO as the weight of the 
$F_L(x,Q^2)$ {\it data} is increased in the fit. 
The best fit results in $\chi^2=27/18$ 
for the $F_L(x,Q^2)$ {\it data}
but this is becoming an unacceptable global fit. 
The next-best fit is an acceptable global fit, and $\chi^2=29/18$ 
for the $F_L(x,Q^2)$ {\it data}. 
The NLO fit to the $F_L(x,Q^2)$ {\it data} is never particularly good because 
the shape in $Q^2$ is never quite correct.
The comparison at NNLO as the weight of the $F_L(x,Q^2)$ {\it data}
is increased in the fit is similar. 
The best fit results in $\chi^2=26/18$ 
for the $F_L(x,Q^2)$ {\it data}
but is becoming an unacceptable global fit. 
The next-best fit is an acceptable global fit, and $\chi^2=31/18$ 
for the $F_L(x,Q^2)$ {\it data}. Again the 
NNLO fit to $F_L(x,Q^2)$ {\it data} always gets the shape in $Q^2$ 
slightly wrong.

%\begin{figure}[ht]
%\centerline{\epsfxsize=4.1in\epsfbox{flnnlocomp.ps}}   
%\caption{Comparison at NNLO as weight of $F_L(x,Q^2)$ {\it data}
%is increased in the fit. 
%\label{flnnlocomp}}
%\end{figure}

\begin{figure}[ht]
\centerline{\epsfxsize=2.4in\epsfbox{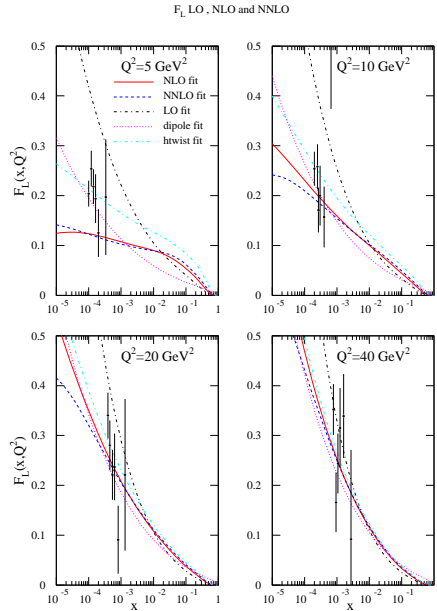}}   
\caption{Comparison of data to predictions with bin at $Q^2=40\GeV^2$ shown.  
\label{flnnloringtwist40}}
\end{figure}

As well as the resummation and dipole hypotheses we  
can also look at explicit higher twist possibilities, in particular the 
renormalon correction due to the nonsinglet quark sector. This is a different
picture from the case for 
$F_2(x,Q^2)$, where the  renormalon calculation of higher twist
dies away at small $x$ due to satisfying the Adler sum rule.
It is a completely different picture for $F_L(x,Q^2)$ -- at small $x$ 
$F_L^{HT}(x,Q^2) \propto F_2(x,Q^2)$.
The explicit renormalon calculation \cite{renormalon} gives
\begin{equation}
F^{HT}_L(x,Q^2) = \frac{A}{Q^2}\otimes F_2(x,Q^2)
\end{equation}
where the estimate for $A$ for the first moment of the structure function is 
\begin{equation}
A= \frac{8C_f\exp(5/3)}{3\beta_0} \Lambda_{QCD}^2
\approx 0.4 \,\GeV^2.
\end{equation}
This effect has nothing to do with the gluon distribution, and is
not part of the higher twist contribution in the dipole approach. 
The higher twist does mix with higher orders though. I add it to 
the NLO prediction. 
The renormalon 
correction could be a rather significant effect, as seen in 
Figure~\ref{flnnloringtwist40}, where I 
generate a new set of data based on the central higher twist 
prediction. (The {\it data} at $Q^2=40\GeV^2$ are shown.  
All predictions give $\chi^2=\sim 6/6$ for the six
points at $Q^2=40\GeV^2$ (except LO)).
It is most similar to the dipole prediction but the data give
$\chi^2 =25/18$ for the dipole prediction curve -- perhaps at the  
edge of distinguishability.
The renormalon-based data are clearly able to rule out the 
central NLO and NNLO curves, but one must  
repeat the study done for {\it dipole data}.

%\begin{figure}[ht]
%\centerline{\epsfxsize=4.1in\epsfbox{flx0001ht.ps}}   
%\caption{Evolution of various predictions for $F_L(x,Q^2)$
%at $x=0.0001$ including higher twist.\label{flx0001ht}}
%\end{figure}

%\begin{figure}[ht]
%\centerline{\epsfxsize=4.1in\epsfbox{flnnloringtwist.ps}}   
%\caption{Comparison of data to predictions.  
%\label{flnnloringtwist}}
%\end{figure}

Again I look at the NLO fit as the weight of the $F_L(x,Q^2)$ {\it data}
is increased in the fit. 
The best fit results in $\chi^2=22/18$ 
for the $F_L(x,Q^2)$ {\it data},
but is an unacceptable global fit -- $\Delta \chi^2 > 60$. 
The next-best fit is a marginally acceptable global fit and 
$\chi^2=27/18$ for the $F_L(x,Q^2)$ {\it data}.
Hence, in this case the NLO fit to $F_L(x,Q^2)$ {\it data} can get 
the shape in $Q^2$ (for $Q^2 \ge 5\GeV^2$) more-or-less right, but
the deterioration in the global fit required to do so 
is worse than for the {\it dipole data}.
The comparison at NNLO as the weight of $F_L(x,Q^2)$ {\it data}
is increased in the fit is again similar to that at NLO. 
The best fit results in $\chi^2=23/18$ 
for $F_L(x,Q^2)$ {\it data}
but is a poor global fit -- $\Delta \chi^2 > 50$. 
The next-best fit is a moderately acceptable global fit, 
and $\chi^2=29/18$ for $F_L(x,Q^2)$ {\it data}.

%\begin{figure}[ht]
%\centerline{\epsfxsize=4.1in\epsfbox{flnlocomptwist.ps}}   
%\caption{Comparison at NLO as weight of $F_L(x,Q^2)$ {\it data}
%is increased in the fit. 
%\label{flnlocomptwist}}
%\end{figure}

%\begin{figure}[ht]
%\centerline{\epsfxsize=4.1in\epsfbox{flnnlocomptwist.ps}}   
%\caption{Comparison at NNLO as weight of $F_L(x,Q^2)$ {\it data}
%is increased in the fit. 
%\label{flnnlocomptwist}}
%\end{figure}

\section{Conclusions}

The measurement of $F_L(x,Q^2)$ seems to be the best way to
determine reliably  
the gluon distribution at low $x$, particularly at low $Q^2$, 
and to determine whether fixed-order 
calculations are sufficient or whether resummations,
or other theoretical extensions may be needed.
Currently we can perform global fits to all up-to-date data 
over a wide range of 
parameter space, and the fit quality is fairly good, but there are 
some minor problems. We could require higher orders, higher twist and/or 
some type of resummation, all of which might have 
a potentially large impact on the predicted $F_L(x,Q^2)$ and other 
quantities. 
Hence, $F_L(x,Q^2)$ is a vital measurement for our understanding of 
precisely how best to 
use perturbative QCD to describe the structure of the 
proton and also for making really reliable predictions and comparisons at 
the LHC. The lowest $Q^2$ possible would be useful. 
The proposed measurement at HERA would  
have a reasonable ability to distinguish between different theoretical 
approaches, due to both the inability to fit $F_L(x,Q^2)$ because of the 
shape and the deterioration in global fits needed 
in order to match the general features of 
$F_L(x,Q^2)$ data, and would play a central role in determining the 
best way to use QCD. 

\section*{Acknowledgments}

I would like to thank Max Klein for supplying me with the simulated 
H1 data for $F_L(x,Q^2)$,  and him and Mandy Cooper-Sarkar, Claire Gwenlan,
Alan Martin and James Stirling for numerous discussions on the subject of the 
longitudinal structure function.

\end{document}